\documentclass[12pt]{article}
\usepackage{graphicx}
\usepackage{amsmath}

\begin{document}

\title{ Ising model versus normal form game }
\author{Serge Galam$^{1}$ and Bernard Walliser $^{2}$ \\
%EndAName
$^1$ Centre de Recherche en \'Epist\'emologie Appliqu\'ee,\\
\'{E}cole Polytechnique, CNRS UMR 7656\\
serge.galam@polytechnique.edu\\
$^{2}$ Paris School of Economics, \\
ENS-EHESS-ENPC, CNRS UMR 8545\\
walliser@mail.enpc.fr\\
}
\date{To appear in Physica A (2010)}
\maketitle

\begin{abstract}
The 2-spin Ising model in statistical mechanics and the 2x2 normal form game
in game theory are compared. All configurations allowed by the second are
recovered by the first when the only concern is about Nash equilibria.\ But
it holds no longer when Pareto optimum condiderations are introduced like in
the prisoner's dilemma. This gap can nevertheless be filled by adding a new
coupling term to the Ising model, even if that term has up to now no
physical meaning. An individual complete bilinear objective function is thus
found to be sufficient to reproduce all possible configurations of a 2x2
game. Using this one-to-one mapping, new perspectives for future research in
both fields can be envisioned.
\end{abstract}

\section{Introduction}

For two decades, statistical mechanics has been more and more used to study
economic systems (Kirman-Zimmermann, 2001; Bourgine-Nadal, 2004) as well as
financial ones (Bhamra, 2000). A well-known example is the application of
statistical mechanics to minority games at the basis of some financial
problems (Coolen, 2005).\ Such a transfer is generally grounded on a raw
analogy, both fields sharing the existence of numerous entities which are
able to be in specific states and interact among themselves. Each entity is
governed by the optimization of an objective function depending on the
states of all of them. The process leads to an overall equilibrium state.
However, if the formal frame is rather similar, the interpretation
underlying the maximization process is very different. While spins are inert
objects which obey objective laws, human beings follow an intentional
behavior. Hence, the analogy is not really substantial, but may nevertheless
be valid at a formal level.

In this paper, two standard models of each discipline are compared, the
Ising model at zero temperature for statistical mechanics and the normal
form with Nash equilibrium for game theory. When both models are restricted
to two entities and two state variables, a detailed comparison becomes
possible. While in statistical mechanics the individual objective function
is an explicit truncated bilinear function in both entities, in game theory
it is a quite general function with no specification. On this basis, the
possible configurations of equilibrium states are computed in each case and
systematically compared.

It appears that any configuration happening in the first case appears as
well in the second.\ However, when one introduces Pareto optimum
considerations like in the prisoner's dilemma, this is no more true.
However, the prisoner's dilemma missing configuration can be recovered by
adding an additive new term to the Ising objective function. This additional
term makes the associated bilinear function complete. It is a linear term
which at present has no physical meaning. Nevertheless, an individual
bilinear objective function which is complete is sufficient to reproduce all
possible configurations of game theory.

The remaining of the paper is organized as follows. The second section
recalls the frameworks used in statistical mechanics and in game theory and
compares their respective assumptions. The third section analyzes the
configurations of equilibrium states obtained in games when using the
restrictive individual objective function of the Ising model. The fourth
section shows that the missing configuration can be recovered by considering
a complete individual bilinear objective function, but can no longer be
obtained from a collective objective function. Some perspectives for new
research in both fields using our mapping are discussed in the last section.

\section{Basic frameworks}

\subsection{Ising model}

In statistical mechanics, the Ising model is most used. Rather powerful to
study magnetic systems, it was proven successful in describing a huge
spectrum of different problems in other fields of physics. Numerous
applications have been also achieved outside physics in particular in
sociophysics (Castellano, Fortunato and Loreto, 2009; Galam, 2008).

The Ising model considers a group of $N$ two-state variables called spins,
where each spin $i=1,...,N$, takes values denoted $s_{i}=\pm 1.\;$These
spins interact by pairs $\{i,j\}$ via two exchange constants $J_{ij}$ and $%
J_{ji}$ whose effect is to produce respectively a local field $%
h_{j\longrightarrow i}=J_{ij}s_{j}$ from spin $j$ on spin $i$ and a local
field $h_{i\longrightarrow j}=J_{ji}s_{i}$ from spin $i$ on spin $j.$These
fields couple linearly with their associated spins giving rise to the two
individual energies:

\ \ \ \ \ \ \ \ \ \ \ \ \ \ \ \ \ \ \ \ \ \ \ \ \ \ \ \ \ \ \ \ \ \ \ \ \ \
\ \ \ \ \ \ \ 
\begin{tabular}{l}
$E_{i}=-h_{j\longrightarrow i}s_{i}=-J_{ij}s_{j}s_{i}$%
\end{tabular}

\ \ \ \ \ \ \ \ \ \ \ \ \ \ \ \ \ \ \ \ \ \ \ \ \ \ \ \ \ \ \ \ \ \ \ \ \ \
\ \ \ \ \ \ \ 
\begin{tabular}{l}
$E_{j}=-h_{i\longrightarrow j}s_{j}=-J_{ji}s_{i}s_{j}$%
\end{tabular}

which are both bilinear functions of the spins.

The minus sign remembers that the energy is ultimately minimized. A positive
coupling $J_{ij}\succ 0$ corresponds to a ferromagnetic system and yields: $%
s_{i}=s_{j}.$ A negative coupling $J_{ij}\prec 0$ corresponds to a
antiferromagnetic system and favors: $s_{i}=-s_{j}.$ The absence of coupling 
$J_{ij}=0$ makes both values of $s_{i}$ and $s_{j}$ independent one from the
other.

Then, considering one spin $s_{i,}$ all its interactions with the other $%
(N-1)$ spins add up to result in a net local field:

\ \ \ \ \ \ \ \ \ \ \ \ \ \ \ \ \ \ \ \ \ \ \ \ \ \ \ \ \ \ \ \ \ \ \ \ \ \
\ \ \ \ \ \ \ 
\begin{tabular}{l}
$V_{i}=\sum_{j=1}^{N}J_{ij}s_{j},$ with $J_{ii}=0$%
\end{tabular}

At this stage, in order to satisfy the principle of equality of action and
reaction which holds true for most physical applications, the exchange
couplings are taken symmetric: $J_{ij}=J_{ji}$

In addition, spins can be coupled linearly to a local field $\tilde{h}_{i}$,
which may vary from spin to spin like in the random field model. It is
independent of the spin states. A uniform field $h$ can also be applied to
the system.\ It couples linearly and simultaneously to every spin. These
coupling terms write respectively $\tilde{h}_{i}s_{i}$ and $hs_{i}$. Both $%
\tilde{h}_{i}$ and $h$ can be positive, negative or null. In physics,
discriminating between $\tilde{h}_{i}$ and $h$ originated from the different
experimental schemes which produce them respectively. On a more formal point
of view, they can be combined within one single local variable: $h_{i}=%
\tilde{h}_{i}+h$.

The total local field becomes:

\ \ \ \ \ \ \ \ \ \ \ \ \ \ \ \ \ \ \ \ \ \ \ \ \ \ \ \ \ \ \ \ \ \ \ \ \ \
\ \ \ \ \ \ \ 
\begin{tabular}{l}
$V_{i}=\sum_{j=1}^{N}J_{ij}s_{j}+h_{i}$%
\end{tabular}

and the associated individual energy:

\ \ \ \ \ \ \ \ \ \ \ \ \ \ \ \ \ \ \ \ \ \ \ \ \ \ \ \ \ \ \ \ \ \ \ \ \ \
\ \ \ \ \ \ \ 
\begin{tabular}{l}
$E_{i}=-\sum_{j=1}^{N}J_{ij}s_{i}s_{j}-h_{i}s_{i}$%
\end{tabular}

It is worth to note that a self-coupling term with $i=j$ may be included in $%
E_{i}$. However since $s_{i}s_{i}=1$, it contributes a constant $J_{ii}$ to $%
E_{i}$, which therefore does not influence the equilibrium state. In the
special case where $V_{i}=0,$ the associated spin $s_{i}$ is said to be
\textquotedblleft frustrated\textquotedblright\ since neither $s_{i}=+1$ nor 
$s_{i}=-1$ can minimize $E_{i}=0$. It makes a spin flip cost no energy.

The total energy of the $N$-spin system is then obtained by adding all
individual energies $E_{i}$ which defines the Hamiltonian under two
equivalent forms: 
\begin{eqnarray}
\mathcal{H} &=&-\sum_{<i,j>}J_{ij}s_{i}s_{j}-\sum_{i}h_{i}s_{i}\text{ \ \ \
\ } \\
\mathcal{H}^{\prime } &=&-\frac{1}{2}\sum_{i,j=1}^{N}J_{ij}s_{i}s_{j}-%
\sum_{i}h_{i}s_{i}
\end{eqnarray}%
where $<i,j>$ means all distinct pairs of interacting spins and $1/2$
corrects the double counting of each pair. In other circumstances, the
system may stabilize into some limit cycle.

The Hamiltonian measures the energy of a peculiar configuration $%
\{s_{1},s_{2},...,s_{N}\}$ given the external parameters $\left\{
J_{ij},h_{i}\right\} $. It is sufficient to describe the equilibrium
properties of the system.

Observe that minimizing the Hamiltonian with respect to the states $\left\{
s_{i}\right\} $ is identical to minimizing each $\left\{ E_{i}\right\} $
with respect to $s_{i}$ . However, it should be emphasized that this feature
is a direct result of the assumption of symmetric couplings ($%
J_{ij}=J_{ji}).\ $Indeed, the term $J_{ij}$ may be virtually asymmetric ($%
J_{ij}\neq J_{ji})$ .\ In that case, the energy function cannot characterize
uniquely the system since then the asymmetric part waves out in summing both
terms $J_{ij}s_{i}s_{j}$ and $J_{ji}s_{j}s_{i}$ in $\mathcal{H}$ due to the
commutation of $s_{i}$ and $s_{j}$ resulting in one average coupling term ($%
J_{ij}+J_{ji})/2$ which is symmetric. Therefore, for asymmetric couplings, a
purely dynamical analysis is required making the solving of the problem much
more difficult.

Observe moreover that, on purely formal grounds, $h_{i}$ can be taken as
positive without restriction. Exchanging the sign of $s_{i}$ preserves the
value of $E_{i}$ if changing altogether the signs of $J_{ij}$ and $h_{i}$.

At this stage, to make the comparison to game theory straightforward, we
restrict the approach to two spins: $N=2,$even if the physical
interpretation is dubious. The local fields becomes:%
\begin{equation}
V_{1}=J_{21}s_{2}+h_{1}\text{ and }V_{2}=J_{12}s_{1}+h_{2}
\end{equation}

The individual energies follow:

\begin{equation}
E_{1}=-J_{21}s_{1}s_{2}-h_{1}s_{1}\text{ and }%
V_{2}=-J_{12}s_{1}s_{2}-h_{2}s_{2}
\end{equation}

Finally, the Hamiltonian reduces to:

\begin{equation}
\mathcal{H}_{1,2}^{\prime }=-\frac{1}{2}%
(J_{12}+J_{21})s_{1}s_{2}-h_{1}s_{1}-h_{2}s_{2}.
\end{equation}

On this basis, the equilibrium state is obtained by a minimization of $%
\mathcal{H}^{\prime }$ with respect to both $\{s_{1},s_{2}\}$ given $%
\{J_{12},J_{21},h_{1},h_{2}\}$. Hence $-\mathcal{H}^{\prime }$ is a
collective objective function which is maximized with respect to the two
internal degrees of freedom$\{s_{1},s_{2}\}$. Remember that $h_{1}$ and $%
h_{2}$ can be taken positive without restriction.

In the case of symmetric exchange couplings ($J_{12}=J_{21}$), the
minimization of $E_{1}$ with respect to $s_{1}$ and $E_{2}$ with respect to $%
s_{2}$ is precisely equivalent to the minimization of $\mathcal{H}$ with
respect to $\{s_{1},s_{2}\}$. The functions $E_{1}$ and $E_{2}$ are
therefore individually minimized, hence their opposite are individual
objective functions which are maximized.

\subsection{Game theory}

In game theory, the standard \textquotedblleft
normal-form\textquotedblright\ is most used.\ It applies to many economic
applications as well as to applications in other fields like political
science or biology.

The framework considers a group of $N$ players numerated with $i=1,...,N$.
Each player $i$ chooses some action $s_{i}$ in a set $S_{i},$which is
defined on a nominal (discrete) scale rather than on a numerical one. The
players interact globally through their actions and reach some common
consequences $C$ for each issue $(s_{1},...s_{i},...s_{N})$ $.$These
consequences are assessed by each player $i$ thanks to a utility or payoff
function $U_{i}=U_{i}(C)=U_{i}(s_{1},...,s_{i},...,s_{N})$ . Such a utility
index is ordinal, hence the utility function, otherwise of any form, is
defined up to a monotone transformation. Finally, each player chooses his
action by maximizing his utility.

Since the maximization implemented by some player depends on what the other
players do, an equilibrium notion has to be defined in order to coordinate
the players' moves. A Nash equilibrium state is a stable state in a fixed
environment, provided that one player only is able to deviate from it at a
time. Technically, a (pure) Nash equilibrium state is an 'issue' (or 'state'
or 'profile of actions' $(s_{1}^{\ast },...,s_{i}^{\ast },...,s_{N}^{\ast })$
where each action is a best response to the others' equilibrium actions: 
\begin{equation}
s_{i}^{\ast }=\arg \max_{s_{i}}U_{i}(s_{1}^{\ast },...,s_{i},...,s_{N}^{\ast
}).
\end{equation}%
In other terms, in an equilibrium state, no player has an incentive to
deviate unilaterally from his action: 
\begin{equation}
U_{i}(s_{1}^{\ast },...s_{i}^{\ast },...s_{N}^{\ast })\geq U_{i}(s_{1}^{\ast
},...s_{i},...s_{N}^{\ast }),\forall s_{i}.
\end{equation}%
A Nash equilibrium is strict if and only if the inequality is strict for
each player. In fact, the equilibrium conditions state that if the players
are placed in an equilibrium state, they stay there.\ But they do not
explain through what dynamical process the players come to an equilibrium
state.

The Nash equilibrium notion assumes that the players follow their own
individual interests (embedded in their utility function which gathers
heterogenous motivations). The Pareto optimum notion considers a collective
point of view (however based on the individual utility functions).\ It
asserts that a state is a Pareto optimum if there exists no other available
state which is better for all players (and strictly better for one of them
at least). Technically, it is defined in two steps. Firstly, an issue $%
(s_{1},...,s_{i},...,s_{N})$ Pareto-dominates another issue $(s_{1}^{\prime
},...,s_{i}^{\prime },...,s_{N}^{\prime })$ if it gives more utility to each
player (and strictly more to one of them)%
\begin{equation}
U_{i}(s_{1},...s_{i},...s_{N})\geq U_{i}(s_{1}^{\prime },...s_{i}^{\prime
},...s_{N}^{\prime }),\forall i,
\end{equation}%
with one strict inequality. Secondly, a Pareto optimum is an issue which is
not Pareto-dominated by any other issue. However, even if a collective order
on global states is considered, no collective objective function is assumed
to exist.

The game is now restricted to two players and two actions per player, i.e. a
2 x 2 game. It is usually represented by a matrix where the lines represent
the two actions $s_{1}^{1}$ and $s_{1}^{2}$ of the first player and the
columns the two actions $s_{2}^{1}$ and $s_{2}^{2}$ of the second player. At
the intersection of an action for each player, the bi-matrix indicates first
the utility of the first player, then the utility of the second player:

$%
\begin{array}{cc}
& \ s_{2}^{1}\text{ \ \ \ \ \ \ \ \ \ \ \ \ \ \ \ \ \ \ \ \ \ \ \ \ \ \ \ \
\ \ \ \ }s_{2}^{2}\text{\ \ \ \ } \\ 
\begin{array}{c}
s_{1}^{1} \\ 
s_{1}^{2}%
\end{array}
& 
\begin{array}[t]{|c|c|}
\hline
(U_{1}(s_{1}^{1},s_{2}^{1}),U_{2}(s_{1}^{1},s_{2}^{1})) & 
(U_{1}(s_{1}^{1},s_{2}^{2}),U_{2}(s_{1}^{1},s_{2}^{2})) \\ \hline
(U_{1}(s_{1}^{2},s_{2}^{1}),U_{2}(s_{1}^{2},s_{2}^{1})) & 
(U_{1}(s_{1}^{2},s_{2}^{2}),U_{2}(s_{1}^{2},s_{2}^{2})) \\ \hline
\end{array}%
\end{array}%
$

\ \ \ \ \ \ \ \ \ \ \ \ \ \ \ \ \ \ \ \ \ \ \ \ \ \ \ \ \ \ \ \ \ \ \ \ \ \
\ \ \ \ \ \ \ \ \ \ \ \ \ \ \ \ \ \ \ \ 

This bi-matrix has indeed quite a general form and can be written with 8
free parameters:

\begin{equation*}
\begin{array}[t]{|c|c|}
\hline
(a,a^{\prime }) & (b,b^{\prime }) \\ \hline
(c,c^{\prime }) & (d,d^{\prime }) \\ \hline
\end{array}%
\end{equation*}%
\newline
\newline
Two classes of games are of special interest. The symmetric games are such
that the two players play symmetric roles and get symmetric utilities: $%
a=a^{\prime },b=c^{\prime },c=b^{\prime },d=d^{\prime }$ . The zero-sum
games are such that the utilities of the players are opposite for each issue
of the game: $a=-a^{\prime },b=-b^{\prime },c=-c^{\prime },d=-d^{\prime }$.
In case of a symmetric game, it exists at least one equilibrium state while
in case of a zero-sum game, it exists at most one equilibrium state.

In the bi-matrix, the (pure) Nash equilibria are obtained as 'wells' of the 
\textit{best response} arrows relating horizontal or vertical issues in the
matrix. The Pareto optima are obtained by eliminating all dominated issues.
The two types of states need not to coincide. For instance, consider the
game with: $a\succ c\succ d\succ b,$ $c^{\prime }=d^{\prime }\succ a^{\prime
}\succ b^{\prime }.$The best responses, the Nash equilibria (a strict Nash
equilibrium is denoted $N$ while a large one is denoted $N^{\prime }$ ) and
the Pareto optima (denoted $P$) are the following :

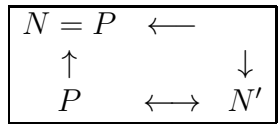
\begin{equation*}
\begin{array}[t]{|ccc|}
\hline
N=P & \longleftarrow &  \\ 
\uparrow &  & \downarrow \\ 
P & \longleftrightarrow & N^{\prime } \\ \hline
\end{array}%
\end{equation*}

Observe that the equilibrium states are unchanged under a monotone
transformation of $U_{1}$ and $U_{2}.$ Especially, they are not affected
when adding to $U_{1}$ (and $U_{2}$) a constant in the first column and
another in the second$:$

\begin{equation*}
\begin{array}[t]{|c|c|}
\hline
(a+\lambda ,a^{\prime }+\lambda ^{\prime }) & (b+\mu ,b^{\prime }+\mu
^{\prime }) \\ \hline
(c+\lambda ,c^{\prime }+\lambda ^{\prime }) & (d+\mu ,d^{\prime }+\mu
^{\prime }) \\ \hline
\end{array}%
\end{equation*}

Likely, the Pareto-dominating issues are unchanged under a linear
transformation of $U_{1}$ and $U_{2}.$ Especially, they are unchanged when
adding to $U_{1}$ (and $U_{2}$) a same constant in all issues:

\begin{equation*}
\begin{array}[t]{|c|c|}
\hline
(a+\lambda ,a^{\prime }+\lambda ^{\prime }) & (b+\lambda ,b^{\prime
}+\lambda ^{\prime }) \\ \hline
(c+\lambda ,c^{\prime }+\lambda ^{\prime }) & (d+\lambda ,d^{\prime
}+\lambda ^{\prime }) \\ \hline
\end{array}%
\end{equation*}

Note however that such transformations do not keep the symmetric or zero-sum
character of the game.

\subsection{Correspondance between the frames}

The two frameworks appear quite similar in their structure: a spin is
replaced by a player, the state $s_{i}$ by an action $s_{i}$ , the local
energy $-E_{i}$ by a utility function $U_{i}$, a physical equilibrium by a
Nash equilibrium. But the Hamiltonian $\mathcal{H}$ has no counterpart since
no collective utility (linear or not) is defined. However, the physical
structure is logically stronger than the game structure since the state is
quantified and the energy is more specific (truncated bilinear function of
the variables).

The physical problem can be transformed into a game problem by constructing
an equivalent normal form for a game. The 2 x 2 matrix gives in each issue
the energy of spins 1 and 2, the issues corresponding to the combinations of
first and second states for the spins. We thus get the 'Ising game' matrix
which is comparable to the one obtained from game theory:

$%
\begin{array}{cc}
& \ \ \ \ \ \ \ \ \ s_{2}^{1}=+1\text{ \ \ \ \ \ \ \ \ \ \ \ \ \ \ \ \ \ \ \
\ \ \ \ \ \ \ \ }s_{2}^{2}=-1\text{\ \ \ \ } \\ 
\begin{array}{c}
s_{1}^{1}=+1 \\ 
s_{1}^{2}=-1%
\end{array}
& 
\begin{array}[t]{|c|c|}
\hline
(J_{12}+h_{1},J_{21}+h_{2}) & (-J_{12}+h_{1},-J_{21}-h_{2}) \\ \hline
(-J_{12}-h_{1},-J_{21}+h_{2}) & (J_{12}-h_{1},J_{21}-h_{2}) \\ \hline
\end{array}%
\end{array}%
$

\ \ \ \ \ \ \ \ \ \ \ \ \ \ \ \ \ \ \ \ \ \ \ 

The same previous general form can be used, but it involves now only 4 free
parameters since the off diagonal payoffs depend on the diagonal payoffs in
the following way:

\begin{equation*}
\begin{array}[t]{|c|c|}
\hline
(\alpha ,\alpha ^{\prime }) & (-\delta ,-\alpha ^{\prime }) \\ \hline
(-\alpha ,-\delta ^{\prime }) & (\delta ,\delta ^{\prime }) \\ \hline
\end{array}%
\end{equation*}

with $\alpha =J_{12}+h_{1}$, $\alpha ^{\prime }=J_{21}+h_{2}$, $\delta
=J_{12}-h_{1}$ and $\delta ^{\prime }=J_{21}-h_{2}$. Note that the
correspondance can be reversed since $J_{12}=(\alpha +\delta
)/2,h_{1}=(\alpha -\delta )/2,J_{21}=(\alpha ^{\prime }+\delta ^{\prime
})/2,h_{2}=(\alpha ^{\prime }-\delta ^{\prime })/2.$

The Ising game is symmetric when $\alpha =\alpha ^{\prime }$ and $\delta
=\delta ^{\prime }$, which involves that: $J_{12}=J_{21}$ and $h_{1}=h_{2}.$
It is zero-sum when $\alpha =\delta =-\alpha ^{\prime }=-\delta ^{\prime },$%
which involves $J_{12}=-J_{21}$ and $h_{1}=h_{2}=0.$

If we are interested only in equilibrium states, it is always possible to
change a usual game into an equivalent Ising game by adding to the utilities
some specific constants:

$\lambda =-(a+c)/2,\mu =-(b+d)/2,\lambda ^{\prime }=-(a^{\prime }+c^{\prime
})/2,\mu ^{\prime }=-(b^{\prime }+d^{\prime })/2$ .

The new utilities become related to the old ones as follows:

$\alpha =(a-c)/2,\delta =(d-b)/2,\alpha ^{\prime }=(a^{\prime }-c^{\prime
})/2,\delta ^{\prime }=(d^{\prime }-b^{\prime })/2.$

Moreover, the physical parameters can be directly related to the old
utilities:

$J_{12}=(a-b-c+d)/2,$ $h_{1}=(a+b-c-d),$

$J_{21}=(a^{\prime }-b^{\prime }-c^{\prime }+d^{\prime })/2,$ $%
h_{2}=(a^{\prime }+b^{\prime }-c^{\prime }-d^{\prime })$

However, if we are interested in Pareto optima, the same operation cannot be
done since the constants generally differ along lines and columns.

\section{\protect\bigskip Comparison of equilibrium states}

\subsection{Taxonomy of 2x2 games}

In general 2x2 games, we are looking for the different types of equilibrium
configurations that may happen. For this, it is enough to consider that the
payoffs of a player are defined up to a monotone transformation, hence are
defined on an ordinal scale. Even more we have only to consider the relative
position of the payoffs leading to the best responses. It means that we have
just to compare $a$ to $c$ and $b$ to $d.$ Such a reasoning concerns in fact
the vertical arrows in the basic scheme. We obtain only 9 combinations:

- 2 combinations A\ without ties and opposite moves:$\uparrow \downarrow $
or $\downarrow \uparrow $

- 2 combinations B\ without ties and parallel moves: $\uparrow \uparrow $ or 
$\downarrow \downarrow $

- 2 combinations C with one tie right:$\uparrow \updownarrow $ or $%
\downarrow \updownarrow $

- 2 combinations D with one tie left:$\updownarrow \uparrow $ or $%
\updownarrow \downarrow $

- 1 combination E with two ties:$\updownarrow \updownarrow $

For two players, we obtain in principle 9x9=81 combinations. But a matrix
obtained by exchanging lines or columns gives the same equilibrium states
since there is only a change of label of the actions. However, for some
matrices, exchanging lines or columns can lead to the same matrix. In fact,
the exchange properties of the different combinations are the following:\ 

- combinations A exchange themselves when exchanging lines and similarly
when exchanging columns.

- combinations B exchange themseves when exchanging lines and are invariant
when exchanging columns

- combinations C exchange themselves when changing lines and lead to
configurations D when exchanging columns

- combinations D exchange themselves when changing lines and lead to
configurations C when exchanging columns

- combination E is invariant when changing lines as well as when changing
columns.

It is clear that only one of the two combinations B, C and D needs to be
kept for each player since the other is obtained by exchanging the lines
(for player 1) and the columns (for player 2). Similarly, the unique
combination E has to be kept. A first problem happens however with
configurations A. Keeping one combination for player 1 is enough when
crossing with any other combination for player 2. But when crossing one
combination A for player 1 and one combination B for player 2 is not enough
since two different configurations are concretely produced. A second problem
appears with configuration E.\ Since it is completely symmetric, its
adoption by player 1 gives the same configuration against the combinations C
and D of player 2, hence one has to be deleted.

The following table makes precise the 15 different configurations that can
be obtained:

\makebox[12cm]{
\begin{tabular}{c|c|c||c|c||c|}
& A & B & C & D & E \\ \hline
$%
\begin{array}{c}
\\ 
\text{A}%
\end{array}%
$ & $%
\begin{array}[t]{ccc}
& \longrightarrow &  \\ 
\uparrow & \left[ \text{1}\right] & \downarrow \\ 
& \longleftarrow & 
\end{array}%
\begin{array}[t]{ccc}
N & \longleftarrow &  \\ 
\uparrow & \left[ \text{2}\right] & \downarrow \\ 
& \longrightarrow & N%
\end{array}%
$ & $%
\begin{array}[t]{ccc}
N & \longleftarrow &  \\ 
\uparrow & \left[ \text{3}\right] & \downarrow \\ 
& \longleftarrow & 
\end{array}%
$ & $%
\begin{array}[t]{ccc}
N & \longleftarrow &  \\ 
\uparrow & \left[ \text{5}\right] & \downarrow \\ 
& \longleftrightarrow & N^{\prime }%
\end{array}%
$ & $%
\begin{array}[t]{ccc}
N^{\prime } & \longleftrightarrow &  \\ 
\uparrow & \left[ \text{8}\right] & \downarrow \\ 
& \longleftarrow & 
\end{array}%
$ & $%
\begin{array}[t]{ccc}
N^{\prime } & \longleftrightarrow &  \\ 
\uparrow & \left[ \text{12}\right] & \downarrow \\ 
& \longleftrightarrow & N^{\prime }%
\end{array}%
$ \\ \hline
$%
\begin{array}{c}
\\ 
\text{B}%
\end{array}%
$ &  & $%
\begin{array}[t]{ccc}
N & \longleftarrow &  \\ 
\uparrow & \left[ \text{4}\right] & \uparrow \\ 
& \longleftarrow & 
\end{array}%
$ & $%
\begin{array}[t]{ccc}
N & \longleftarrow &  \\ 
\uparrow & \left[ \text{6}\right] & \uparrow \\ 
& \longleftrightarrow & 
\end{array}%
$ & $%
\begin{array}[t]{ccc}
N^{\prime } & \longleftrightarrow & N^{\prime } \\ 
\uparrow & \left[ \text{9}\right] & \uparrow \\ 
& \longleftarrow & 
\end{array}%
$ & $%
\begin{array}[t]{ccc}
N^{\prime } & \longleftrightarrow & N^{\prime } \\ 
\uparrow & \left[ \text{13}\right] & \uparrow \\ 
& \longleftrightarrow & 
\end{array}%
$ \\ \hline\hline
$%
\begin{array}{c}
\\ 
\text{C}%
\end{array}%
$ &  &  & $%
\begin{array}[t]{ccc}
N & \longleftarrow &  \\ 
\uparrow & \left[ \text{7}\right] & \updownarrow \\ 
& \longleftrightarrow & N^{\prime }%
\end{array}%
$ & $%
\begin{array}[t]{ccc}
N^{\prime } & \longleftrightarrow & N^{\prime } \\ 
\uparrow & \left[ \text{10}\right] & \updownarrow \\ 
& \longleftarrow & 
\end{array}%
$ & $%
\begin{array}[t]{ccc}
N^{\prime } & \longleftrightarrow & N^{\prime } \\ 
\uparrow & \left[ \text{14}\right] & \updownarrow \\ 
& \longleftrightarrow & N^{\prime }%
\end{array}%
$ \\ \hline
$%
\begin{array}{c}
\\ 
\text{D}%
\end{array}%
$ &  &  &  & $%
\begin{array}[t]{ccc}
N^{\prime } & \longleftrightarrow & N^{\prime } \\ 
\updownarrow & \left[ \text{11}\right] & \uparrow \\ 
N^{\prime } & \longleftarrow & 
\end{array}%
$ &  \\ \hline\hline
$%
\begin{array}{c}
\\ 
\text{E}%
\end{array}%
$ &  &  &  &  & $%
\begin{array}[t]{ccc}
N^{\prime } & \longleftrightarrow & N^{\prime } \\ 
\updownarrow & \left[ \text{15}\right] & \updownarrow \\ 
N^{\prime } & \longleftrightarrow & N^{\prime }%
\end{array}%
$ \\ \hline
\end{tabular}
\newline
\newline
}

The four configurations without ties are the basic ones and can each can be
illustrated by a specific example:

- configuration 1 corresponds to the absence of a Nash equilibrium state.\
It is illustrated by \textquotedblleft matching pennies\textquotedblright\
where both players exhibit one of the two sides of a personal penny; the
first wins if the two pennies are on the same side while the second wins if
they are on opposite sides\newline
\ \ \ \ \ \ \ \ \ \ \ \ \ \ \ \ \ \ \ \ \ \ \ \ \ \ \ 
\begin{tabular}{|l|l|}
\hline
$(1,-1)$ & $(-1,1)$ \\ \hline
$(-1,1)$ & $(1,-1)$ \\ \hline
\end{tabular}

- configuration 2 corresponds to two Nash equilibrium states. A first
subcase happens when the two equilibria cannot be compared: one is better
for the first player and the other is better for the second player. It is
illustrated by the \textquotedblleft battle of the sexes\textquotedblright\
where player 1 is the husband and player 2 the wife; for the husband, the
first action consists in going to boxing and the second to ballet; for the
wife, the first action consists in going to ballet and the second to boxing;
the utility of preferred entertainment is 2 and the utility of being
together to 4.\ A second subcase happens when one equilibrium Pareto
dominates the other. It is illustrated by the \textquotedblleft stag hunt
game\textquotedblright\ where for each player, the first action consists in
hunting a rabbit and the second a deer; the rabbit can be hunt alone and
provides then a utility of 1; the deer can only be hunt conjointly and
provides a utility of 7 for each; if only one player hunts the deer, he will
starve and his utility is -1.

\ \ \ \ \ \ \ \ \ \ \ \ \ \ \ \ \ \ \ \ \ 
\begin{tabular}{|l|l|}
\hline
$(2,2)$ & $(6,4)$ \\ \hline
$(4,6)$ & $(0,0)$ \\ \hline
\end{tabular}
\ \ \ \ \ \ \ \ \ \ \ \ \ \ \ \ \ \ \ \ \ \ \ \ \ \ \ \ \ \ \ \ \ \ \ \ \ \
\ \ \ \ \ 
\begin{tabular}{|l|l|}
\hline
$(7,7)$ & $(-1,1)$ \\ \hline
$(1,-1)$ & $(1,1)$ \\ \hline
\end{tabular}

- configuration 4 corresponds to one Nash equilibrium obtained by dominant
actions for each player. A first subcase happens when the Nash equilibrium
is not Pareto dominated by another issue. It is illustrated by the "lonely
game" where each player buys a toy independently from the other's action;
the utility of the toy for each player is +1 and the desutility of not
buying is -1.\ A second subcase happens when the Nash equilibrium is Pareto
dominated by another issue. It is illustrated by the "prisoner's
dilemma"where for each player, the first action consists in defecting and
the second in cooperating; they get utility 2 if both defect, utility 4 if
both cooperate, but if one cooperates without the other, he receives 0 and
the other receives 6; the only equilibrium state is the issue (2,2); but the
last is not a Pareto optimum (contrary to all other three issues) since it
is Pareto-dominated by the opposite issue (4,4)\newline

\ \ \ \ \ \ \ \ \ \ \ \ \ \ \ \ \ \ \ \ \ \ 
\begin{tabular}{|l|l|}
\hline
$(1,1)$ & $(1,-1)$ \\ \hline
$(-1,1)$ & $(-1,-1)$ \\ \hline
\end{tabular}
\ \ \ \ \ \ \ \ \ \ \ \ \ \ \ \ \ \ \ \ \ \ \ \ \ \ \ \ \ \ \ \ \ \ \ \ \ 
\begin{tabular}{|l|l|}
\hline
$(2,2)$ & $(6,0)$ \\ \hline
$(0,6)$ & $(4,4)$ \\ \hline
\end{tabular}%
\newline
\newline

\subsection{Taxonomy of Ising games}

The Nash equilibrium states can be computed and the configurations they
determine depend only on the relation between $J_{ij}$ and $h_{i}$. The
matrix defined by these parameters isolate 25 basic configurations. For each
configuration, the scheme of best responses in the 2 x 2 matrix is defined
in the following table in which $h_{1}$and $h_{2}$ are strictly positive:

\makebox[12cm]{
\begin{tabular}{c|c|c||c|c||c|}
& $J_{21}<-h_{2}$ & $J_{21}=-h_{2}$ & $-h_{2}<J_{21}<h_{2}$ & $J_{21}=h_{2}$
& $J_{21}>h_{2}$ \\ \hline
$%
\begin{array}{c}
\\ 
J_{12}<-h_{1}%
\end{array}%
$ & $%
\begin{array}[t]{ccc}
& \longrightarrow & N \\ 
\downarrow & \left[ \text{2}\right] & \uparrow \\ 
N & \longleftarrow & 
\end{array}%
$ & $%
\begin{array}[t]{ccc}
& \longleftrightarrow & N^{\prime } \\ 
\downarrow & \left[ \text{5}\right] & \uparrow \\ 
N & \longleftarrow & 
\end{array}%
$ & $%
\begin{array}[t]{ccc}
& \longleftarrow &  \\ 
\downarrow & \left[ \text{3}\right] & \uparrow \\ 
N & \longleftarrow & 
\end{array}%
$ & $%
\begin{array}[t]{ccc}
& \longleftarrow &  \\ 
\downarrow & \left[ \text{8}\right] & \uparrow \\ 
N^{\prime } & \longleftrightarrow & 
\end{array}%
$ & $%
\begin{array}[t]{ccc}
& \longleftarrow &  \\ 
\downarrow & \left[ \text{1}\right] & \uparrow \\ 
& \longrightarrow & 
\end{array}%
$ \\ \hline
$%
\begin{array}{c}
\\ 
J_{12}=-h_{1}%
\end{array}%
$ & $%
\begin{array}[t]{ccc}
& \longrightarrow & N \\ 
\updownarrow & \left[ \text{5}\right] & \uparrow \\ 
N^{\prime } & \longleftarrow & 
\end{array}%
$ & $%
\begin{array}[t]{ccc}
N^{\prime } & \longleftrightarrow & N^{\prime } \\ 
\updownarrow & \left[ \text{11}\right] & \uparrow \\ 
N^{\prime } & \longleftarrow & 
\end{array}%
$ & $%
\begin{array}[t]{ccc}
N^{\prime } & \longleftarrow &  \\ 
\updownarrow & \left[ \text{9}\right] & \uparrow \\ 
N^{\prime } & \longleftarrow & 
\end{array}%
$ & $%
\begin{array}[t]{ccc}
N^{\prime } & \longleftarrow &  \\ 
\updownarrow & \left[ \text{10}\right] & \uparrow \\ 
N^{\prime } & \longleftrightarrow & 
\end{array}%
$ & $%
\begin{array}[t]{ccc}
N^{\prime } & \longleftarrow &  \\ 
\updownarrow & \left[ \text{8}\right] & \uparrow \\ 
& \longrightarrow & 
\end{array}%
$ \\ \hline\hline
$%
\begin{array}{c}
\\ 
-h_{1}<J_{12} \\ 
J_{12}<h_{1}%
\end{array}%
$ & $%
\begin{array}[t]{ccc}
& \longrightarrow & N \\ 
\uparrow & \left[ \text{3}\right] & \uparrow \\ 
& \longleftarrow & 
\end{array}%
$ & $%
\begin{array}[t]{ccc}
N^{\prime } & \longleftrightarrow & N^{\prime } \\ 
\uparrow & \left[ \text{9}\right] & \uparrow \\ 
& \longleftarrow & 
\end{array}%
$ & $%
\begin{array}[t]{ccc}
N & \longleftarrow &  \\ 
\uparrow & \left[ \text{4}\right] & \uparrow \\ 
& \longleftarrow & 
\end{array}%
$ & $%
\begin{array}[t]{ccc}
N & \longleftarrow &  \\ 
\uparrow & \left[ \text{6}\right] & \uparrow \\ 
& \longleftrightarrow & 
\end{array}%
$ & $%
\begin{array}[t]{ccc}
N & \longleftarrow &  \\ 
\uparrow & \left[ \text{3}\right] & \uparrow \\ 
& \longrightarrow & 
\end{array}%
$ \\ \hline
$%
\begin{array}{c}
\\ 
J_{12}=h_{1}%
\end{array}%
$ & $%
\begin{array}[t]{ccc}
& \longrightarrow & N^{\prime } \\ 
\uparrow & \left[ \text{8}\right] & \updownarrow \\ 
& \longleftarrow & 
\end{array}%
$ & $%
\begin{array}[t]{ccc}
N^{\prime } & \longleftrightarrow & N^{\prime } \\ 
\uparrow & \left[ \text{10}\right] & \updownarrow \\ 
& \longleftarrow & 
\end{array}%
$ & $%
\begin{array}[t]{ccc}
N & \longleftarrow &  \\ 
\uparrow & \left[ \text{6}\right] & \updownarrow \\ 
& \longleftarrow & 
\end{array}%
$ & $%
\begin{array}[t]{ccc}
N & \longleftarrow &  \\ 
\uparrow & \left[ \text{7}\right] & \updownarrow \\ 
& \longleftrightarrow & N^{\prime }%
\end{array}%
$ & $%
\begin{array}[t]{ccc}
N & \longleftarrow &  \\ 
\uparrow & \left[ \text{5}\right] & \updownarrow \\ 
& \longrightarrow & N^{\prime }%
\end{array}%
$ \\ \hline\hline
$%
\begin{array}{c}
\\ 
J_{12}>h_{1}%
\end{array}%
$ & $%
\begin{array}[t]{ccc}
& \longrightarrow &  \\ 
\uparrow & \left[ \text{1}\right] & \downarrow \\ 
& \longleftarrow & 
\end{array}%
$ & $%
\begin{array}[t]{ccc}
N^{\prime } & \longleftrightarrow &  \\ 
\uparrow & \left[ \text{8}\right] & \downarrow \\ 
& \longleftarrow & 
\end{array}%
$ & $%
\begin{array}[t]{ccc}
N & \longleftarrow &  \\ 
\uparrow & \left[ \text{3}\right] & \downarrow \\ 
& \longleftarrow & 
\end{array}%
$ & $%
\begin{array}[t]{ccc}
N & \longleftarrow &  \\ 
\uparrow & \left[ \text{5}\right] & \downarrow \\ 
& \longleftrightarrow & N^{\prime }%
\end{array}%
$ & $%
\begin{array}[t]{ccc}
N & \longleftarrow &  \\ 
\uparrow & \left[ \text{2}\right] & \downarrow \\ 
& \longrightarrow & N%
\end{array}%
$ \\ \hline
\end{tabular}
}
\newline
\newline
The preceding table has to be completed by a table where $h_{1}=0,h_{2}>0$
(or symetrically):

\makebox[12cm]{
\begin{tabular}{c|c|c||c|c||c|}
& $J_{21}<-h_{2}$ & $J_{21}=-h_{2}$ & $-h_{2}<J_{21}<h_{2}$ & $J_{21}=h_{2}$
& $J_{21}>h_{2}$ \\ \hline
$%
\begin{array}{c}
\\ 
J_{12}<0%
\end{array}%
$ & $%
\begin{array}[t]{ccc}
& \longrightarrow & N \\ 
\downarrow & \left[ \text{2}\right] & \uparrow \\ 
N & \longleftarrow & 
\end{array}%
$ & $%
\begin{array}[t]{ccc}
& \longleftrightarrow & N^{\prime } \\ 
\downarrow & \left[ \text{5}\right] & \uparrow \\ 
N & \longleftarrow & 
\end{array}%
$ & $%
\begin{array}[t]{ccc}
& \longleftarrow &  \\ 
\downarrow & \left[ \text{3}\right] & \uparrow \\ 
N & \longleftarrow & 
\end{array}%
$ & $%
\begin{array}[t]{ccc}
& \longleftarrow &  \\ 
\downarrow & \left[ \text{8}\right] & \uparrow \\ 
N & \longleftrightarrow & 
\end{array}%
$ & $%
\begin{array}[t]{ccc}
& \longleftarrow &  \\ 
\downarrow & \left[ \text{1}\right] & \uparrow \\ 
& \longrightarrow & 
\end{array}%
$ \\ \hline
$%
\begin{array}{c}
\\ 
J_{12}=0%
\end{array}%
$ & $%
\begin{array}[t]{ccc}
& \longrightarrow & N^{\prime } \\ 
\updownarrow & \left[ \text{12}\right] & \updownarrow \\ 
N^{\prime } & \longleftarrow & 
\end{array}%
$ & $%
\begin{array}[t]{ccc}
N^{\prime } & \longleftrightarrow & N^{\prime } \\ 
\updownarrow & \left[ \text{14}\right] & \updownarrow \\ 
N^{\prime } & \longleftarrow & 
\end{array}%
$ & $%
\begin{array}[t]{ccc}
N^{\prime } & \longleftarrow &  \\ 
\updownarrow & \left[ \text{13}\right] & \updownarrow \\ 
N^{\prime } & \longleftarrow & 
\end{array}%
$ & $%
\begin{array}[t]{ccc}
N^{\prime } & \longleftarrow &  \\ 
\updownarrow & \left[ \text{14}\right] & \updownarrow \\ 
N^{\prime } & \longleftrightarrow & N^{\prime }%
\end{array}%
$ & $%
\begin{array}[t]{ccc}
N & \longleftarrow &  \\ 
\updownarrow & \left[ \text{12}\right] & \updownarrow \\ 
& \longrightarrow & 
\end{array}%
$ \\ \hline\hline
$%
\begin{array}{c}
\\ 
J_{12}>0%
\end{array}%
$ & $%
\begin{array}[t]{ccc}
& \longrightarrow &  \\ 
\uparrow & \left[ \text{1}\right] & \downarrow \\ 
& \longleftarrow & 
\end{array}%
$ & $%
\begin{array}[t]{ccc}
N & \longleftrightarrow &  \\ 
\uparrow & \left[ \text{8}\right] & \downarrow \\ 
& \longleftarrow & 
\end{array}%
$ & $%
\begin{array}[t]{ccc}
N & \longleftarrow &  \\ 
\uparrow & \left[ \text{3}\right] & \downarrow \\ 
& \longleftarrow & 
\end{array}%
$ & $%
\begin{array}[t]{ccc}
N & \longleftarrow &  \\ 
\uparrow & \left[ \text{5}\right] & \downarrow \\ 
& \longleftrightarrow & N^{\prime }%
\end{array}%
$ & $%
\begin{array}[t]{ccc}
N & \longleftarrow &  \\ 
\uparrow & \left[ \text{2}\right] & \downarrow \\ 
& \longrightarrow & N%
\end{array}%
$ \\ \hline
\end{tabular}
}
\newline

Finally, a last table is obtained when $h_{1}=h_{2}=0:$\ \ \ \ \ \ \ \ \ \ \
\ \ \ \ \ \ \ \ \ \ \ \ \ \ \

\begin{tabular}{c|c|c||c|}
& $J_{21}<0$ & $J_{21}=0$ & $J_{21}>0$ \\ \hline
$%
\begin{array}{c}
\\ 
J_{12}<0%
\end{array}%
$ & $%
\begin{array}[t]{ccc}
& \longrightarrow & N^{\prime } \\ 
\downarrow & \left[ \text{2}\right] & \uparrow \\ 
N^{\prime } & \longleftarrow & 
\end{array}%
$ & $%
\begin{array}[t]{ccc}
& \longleftrightarrow & N^{\prime } \\ 
\downarrow & \left[ \text{12}\right] & \uparrow \\ 
N^{\prime } & \longleftrightarrow & 
\end{array}%
$ & $%
\begin{array}[t]{ccc}
& \longleftarrow &  \\ 
\downarrow & \left[ \text{1}\right] & \uparrow \\ 
& \longrightarrow & 
\end{array}%
$ \\ \hline
$%
\begin{array}{c}
\\ 
J_{12}=0%
\end{array}%
$ & $%
\begin{array}[t]{ccc}
& \longrightarrow & N^{\prime } \\ 
\updownarrow & \left[ \text{12}\right] & \updownarrow \\ 
N^{\prime } & \longleftarrow & 
\end{array}%
$ & $%
\begin{array}[t]{ccc}
N^{\prime } & \longleftrightarrow & N^{\prime } \\ 
\updownarrow & \left[ \text{15}\right] & \updownarrow \\ 
N^{\prime } & \longleftrightarrow & N^{\prime }%
\end{array}%
$ & $%
\begin{array}[t]{ccc}
N^{\prime } & \longleftarrow &  \\ 
\updownarrow & \left[ \text{12}\right] & \updownarrow \\ 
& \longrightarrow & N^{\prime }%
\end{array}%
$ \\ \hline\hline
$%
\begin{array}{c}
\\ 
J_{12}>0%
\end{array}%
$ & $%
\begin{array}[t]{ccc}
& \longrightarrow &  \\ 
\uparrow & \left[ \text{1}\right] & \downarrow \\ 
& \longleftarrow & 
\end{array}%
$ & $%
\begin{array}[t]{ccc}
N^{\prime } & \longleftrightarrow &  \\ 
\uparrow & \left[ \text{12}\right] & \downarrow \\ 
& \longleftrightarrow & N^{\prime }%
\end{array}%
$ & $%
\begin{array}[t]{ccc}
N^{\prime } & \longleftarrow &  \\ 
\uparrow & \left[ \text{2}\right] & \downarrow \\ 
& \longrightarrow & N^{\prime }%
\end{array}%
$ \\ \hline
\end{tabular}%

Of course, many configurations appear twice by symmetry. But more important,
all configurations appear as expected. The first table gives configurations
1 to 11.\ The second table adds the configurations 12 to 14. The third table
adds the configuration 15.

Notice that the game may be symmetric only in configurations 2, 4, 7 ,11 and
15. It can be checked that it then admits at least one equilibrium state.
Likely, the game may be zero-sum only in configurations 1 and 15. It then
admits either zero or four Nash equilibrium states $.$

The games without ties can be obtained from Ising models for specific
parameters:

- configuration 1 (no equilibrium state) is obtained when $J_{ij}J_{ji}<0$
and $\mid J_{ij}\mid >\mid h_{i}\mid .$Especially, matching pennies is
obtained with $J_{12}=$ \ $1,$ $J_{21}$ \ $=-1$ and $\ h_{1}$ $=$ $h_{2}=0.$

- configuration 2 (two strict equilibrium states) is obtained when $%
J_{ij}J_{ji}>0$ and $\mid J_{ij}\mid >\mid h_{i}\mid $ .\ Especially, the
battle of sexes is obtained with $J_{12}=$ \ $J_{21}$ \ $=-2$ and $\ h_{1}$ $%
=$ $h_{2}=1$ and a general translation of utility of $3.$The stag hunt game
is obtained with $J_{12}=$ \ $J_{21}$ \ $=2$ and $h_{1}$ $=$ $h_{2}=+1$ and
a general translation of utility of $4$

- configuration 4 (one strict equilibrium state) is obtained when $\mid
J_{ij}\mid <\mid h_{i}\mid .\ $Especially, the lonely game is obtained for $%
J_{12}=J_{21}=0$ and $h_{1}=h_{2}=1$

.

\section{Extended framework}

\subsection{ Missing configurations}

In the preceding framework, only the point of view of Nash equilibrium
states was considered. A Nash equilibrium corresponds to a state where the
players may stay, when they are astrained to strictly follow their own
interest. But the game theorists are interested in Pareto optima too
(especially when they coincide with equilibrium states). The reason is that
an issue which is not Pareto optimal is not collectively satisfying since
the players may move to a better issue for both, however by coordinating
their actions.\ 

Consider an Ising game where the North-West issue $(\alpha ,\alpha ^{\prime
})$ is an equilibrium and the South-West issue $(\delta ,\delta ^{\prime })$%
\ is not. The corresponding conditions are $\alpha \succ 0$ , $\alpha
^{\prime }\succ 0^{{}}$ and $\delta \prec 0,\ \delta ^{\prime }\prec 0.\ $It
follows immediatly that the North-West issue $(\alpha ,\alpha ^{\prime })$
Pareto-dominates the South-West one $(\delta ,\delta ^{\prime })$. The
opposite dominance is not possible, hence some games which precisely show
such an opposite dominance cannot be represented by Ising games.

For instance, consider the "prisoner's dilemma". The North-West issue (where
both players defect) is the only equilibrium state .\ The South-West issue
(where both players cooperate) is better for both players to the opposite
one. Hence, the prisoner's dilemma, with its specific structure of a unique
Nash equilibrium Pareto-dominated by a non equilibrium issue, cannot be
written as an Ising game.

\subsection{\ \protect\bigskip Enlarged framework}

In order to recover the missing configuration within the Ising model and
thus get a full matching of configurations, the simplest way is to add to
the local energy of each spin an 'altruistic' term $k_{i}s_{j}$, which
depends exclusively on the other spin state: 
\begin{equation}
-E_{i}=-\sum_{j}J_{ij}s_{i}s_{j}-h_{i}s_{i}-k_{i}s_{j}.
\end{equation}

In fact, such an operation transforms, for player $i$, his initial payoff in
a payoff where $k_{i}$ is added in the first column and where $k_{i}$ is
substracted in the second column. Such an operation does not change the
equilibrium states (even if it changes the payoffs at equilibrium), but does
change the Pareto states. The 'extended Ising game' writes now in the form
of a general game:

$%
\begin{array}{cc}
& \ \ \ \ \ \ \ \ \ s_{2}^{1}=+1\text{ \ \ \ \ \ \ \ \ \ \ \ \ \ \ \ \ \ \ \
\ \ \ \ \ \ \ \ }s_{2}^{2}=-1\text{\ \ \ \ } \\ 
\begin{array}{c}
s_{1}^{1}=+1 \\ 
s_{1}^{2}=-1%
\end{array}
& 
\begin{array}[t]{|c|c|}
\hline
(J_{12}+h_{1}+k_{1},J_{21}+h_{2}+k_{2}) & 
(-J_{12}+h_{1}-k_{1},-J_{21}-h_{2}+k_{2}) \\ \hline
(-J_{12}-h_{1}+k_{1},-J_{21}+h_{2}-k_{2}) & 
(J_{12}-h_{1}-k_{1},J_{21}-h_{2}-k_{2}) \\ \hline
\end{array}%
\end{array}%
$

This matrix is written under the form:

\begin{equation*}
\begin{array}[t]{|c|c|}
\hline
(\overline{a},\overline{a}^{\prime }) & (\overline{b},\overline{b}^{\prime })
\\ \hline
(\overline{c},\overline{c}^{\prime }) & (\overline{d},\overline{d}^{\prime })
\\ \hline
\end{array}%
\end{equation*}%
\newline

with $\overline{a}=J_{12}+h_{1}+k_{1},\overline{b}=$ \ $-J_{12}+h_{1}-k_{1}$%
\ $,\overline{c}=$\ $-J_{12}-h_{1}+k_{1}$ and $\overline{d}=$\ \ $%
J_{12}-h_{1}-k_{1}$\ for player 1 and similarly for player 2.\ However, the
coefficients of the first player still obey the constraint: $\overline{a}+%
\overline{b}+\overline{c}+\overline{d}=0$ (and symmetrically for the
second). But this condition can always be reached from the original matrix
by substracting to the coefficients of the first player a constant $%
(a+b+c+d)/4$ to all payoffs (without changing the equilibrium and Pareto
states). The preceding formulas can then be reversed (for the first player):

$J_{12}=(\overline{a}-\overline{b}-\overline{c}+\overline{d})/4=(a-b-c+d)/4$

$h_{1}=(\overline{a}+\overline{b}-\overline{c}-\overline{d})/4=(a+b-c-d)/4$

$k_{1}=(\overline{a}-\overline{b}+\overline{c}-\overline{d})/4=(a-b+c-d)/4%
\newline
$Hence, a complete bilinear function is sufficient to obtain all possible
configurations.

For instance, the prisoner's dilemma is obtained by setting $J_{12}=J_{21}=0$
, $h_{1}=h_{2}=1$ and $k_{1}=k_{2}=-2,$ and applying a general payoff
transformation of $+3.$\newline
\newline

\subsection{The 'altruistic' term}

At this stage, two comments can be made about the 'altruistic term' added to
the energy of each spin. First, this term has no clear interpretation in
statistical mechanics. Second, the completed energy cannot be deduced from a
collective single Hamiltonian.

If such an Hamiltonian existed, he would be obtained by 'integrating' the
new energy of each spin $V_{i}$, obtained by divising $-E_{i}$ by $s_{i}$ :

\begin{eqnarray}
V_{1}&=&\frac{\partial \mathcal{H}}{\partial s_{1}}
=J_{12}s_{2}+h_{1}+k_{1}s_{2}/s_{1} \\
V_{2}&=&\frac{\partial \mathcal{H}}{\partial s_{2}}
=J_{21}s_{1}+h_{2}+k_{2}s_{1}/s_{2}
\end{eqnarray}

But since the second derivatives $\frac{\partial ^{2}\mathcal{H}}{\partial
s_{1}\partial s_{2}}$and $\frac{\partial ^{2}\mathcal{H}}{\partial
s_{2}\partial s_{1}}$\ differ, the Hamiltonian does not exist.

Conversely, consider the Hamiltonian obtained by adding the energy of both
spins.

\begin{equation}
\mathcal{H}^{\prime
}=J_{12}s_{1}s_{2}+J_{21}s_{1}s_{2}+h_{1}s_{1}+h_{2}s_{2}+k_{1}s_{2}+k_{2}s_{1}
\end{equation}

It leads to individual spin energies which are not the good ones, but it
allows to find the Pareto optimum in the prisoner's dilemma. Hence, the
missing configuration, contrary to the other ones, cannot be obtained by
minimizing a Hamiltonian.

\section{Further extensions}

Extension to continuous positions for each entity is usual in games as well
as in statistical mechanics and their comparison can be extended. But
extension to N entities creates a further divergence between statistical
mechanics and game theory. In statistical mechanics, the energy of a spin is
generally generated by summarizing its bilateral interactions with all other
spins. Multilateral interactions are possible in some cases, but are more
difficult to handle mathematically. In economics, the utility is directly
defined on a multilateral basis, by a general dependence on all others'
actions. Again, it is not specified, but only submitted to very general
conditions. It follows that the additional term(s) needed to recover a
general function from a partial one will be intricate.

Two problems stay common to statistical mechanics and game theory. In both
cases, the equilibrium state is defined from outside (by the modeler) as a
state which remains stable without perturbations from the environment. But
the concrete process leading to some well-defined equilibrium state is not
formally described. Hence, the '\textit{implementation problem}' states that
the process leading to an equilibrium state has to be made explicit. Likely,
the '\textit{selection problem}' states that the way one equilibrium state
is sorted out in case of multiplicity has to be made explicit, in relation
with the preceding one. In physics, a random selection is often assumed with
equiprobability.

In game theory, two justifications of an equilibrium notion were recently
proposed and duly formalized (Walliser, 2006). The epistemic justifications
assume that the players are hyper-intelligent and are able to reach
instantaneously an equilibrium state by their sole reasoning. Moreover, the
players select one equilibrium state through shared 'conventions', for
instance focal points (culturally induced). The evolutionist justifications
assume that the players have a bounded rationality and follow a sequential
learning or evolution process leading asymptotically to an equilibrium
state. In that case, an equilibrium state is automatically selected at least
probabilistically by the initial conditions and the history of the system.

In statistical mechanics, only the second option is open since the spins can
hardly be assumed to hold beliefs and to deliberate. Nevertheless the
sociophysics approach has proven possible to substitute individuals to spins
as long as the assumptions made are formally equivalent. In fact, various
schemes were proposed in order to describe the local dynamic genesis of an
equilibrium state, with more or less justifications. They allow the
convergence towards an equilibrium state by using the concept of spontaneous
symmetry breaking.

Finally, statistical mechanics studies the system of spins even when the
temperature differs from zero. A spin is then in some state with a
probability precisely defined which decreases with its energy and increases
with the inverse of the temperature. In game theory, the corresponding
situation happens when a player chooses an action with a probability
generically defined which increases with its utility and depends on some
parameter characterizing the 'exploration- exploitation' dilemma. The
corresponding equilibrium states have to be further compared.

\section{\protect\bigskip References}

\bigskip

Arthur, W.B.- Durlauf, S. - Lane, D. (1997): \textit{The economy as a
complex evolving system II}, Perseus Books.

Bhamra, H.S. (2000): Imitation in financial markets, \textit{International
Journal of Theoretical and Applied Finance}, 3(3), 473-78.

Blume, M.(1993): The statistical mechanics of strategic interaction, \textit{%
Games and Economic Behavior}, 5, 387-424.

Bourgine, P.- Nadal, J.P. eds.(2004): \textit{Cognitive economics}, Springer.

Castellano, C.-Fortunato, S.-Loreto, V. (2009): Statistical physics of
modern dynamics, Rev. Mod. Phys. 81, 591-646 (2009)

Coolen, A.C.C. (2005): \textit{The mathematical theory of minority games:
statistical mechanics of interacting agents}, Oxford University Press.

Galam, S. (2008): Sociophysics: a review of Galam models, \textit{%
International Journal of Modern Physics} C, 19(3), 409-440

Kirman, A.- Zimmermann, J.B. eds. (2001): \textit{Economics with
heterogenous interacting agents}, Springer.

Walliser, B. (2006): Justifications of game equilibrium notions, in R.\
Arena, A.\ Festre eds: \textit{Knowledge, beliefs and economics}, Edward
Elgar.

\bigskip

\end{document}